\documentclass[letterpaper]{article}

\usepackage{hyperref}
\usepackage[super]{nth}
\usepackage[gen]{eurosym}
\usepackage{bbm}
\usepackage{mathtools}
\usepackage{tikz}
\usepackage{cite}
\usepackage{amsmath,amssymb,amsfonts}
\usepackage[noend]{algorithmic}
\algsetup{indent=3ex}

\usepackage{graphicx}
\usepackage{textcomp}
\usepackage{xcolor}
\usepackage{algorithm}
\usepackage{url}
\usepackage{booktabs}
\usepackage{placeins}
\usepackage{csquotes}
\usepackage[group-separator={,}]{siunitx}
\usepackage{paralist}
\usepackage{listings}
\usepackage{xspace}
\usepackage{appendix}

\newcommand{\lnew}{\ell^{\text{new}}}
\newcommand{\unew}{u^{\text{new}}}
\newcommand{\code}[1]{\texttt{#1}}

\def\BibTeX{{\rm B\kern-.05em{\sc i\kern-.025em b}\kern-.08em
    T\kern-.1667em\lower.7ex\hbox{E}\kern-.125emX}}

\newtheorem{definition}{Definition}

\DeclarePairedDelimiterX{\normSimple}[1]{\lVert}{\rVert}
{\ifblank{#1}{\mathord{\cdot}}{#1}}

\hypersetup{
    colorlinks,
    linkcolor={blue!50!black},
    citecolor={blue!50!black}
}

\begin{document}

\begin{center}
 {\LARGE Accelerating Domain Propagation: \\an Efficient GPU-Parallel Algorithm over Sparse Matrices}
\end{center}

\vspace{8mm}

\textbf{Boro Sofranac}\hfill{\ttfamily sofranac@zib.de}\\
{\small\emph{Berlin Institute of Technology and Zuse Institute Berlin}}

\textbf{Ambros Gleixner}\hfill{\ttfamily gleixner@zib.de}\\
{\small\emph{HTW Berlin and Zuse Institute Berlin}}

\textbf{Sebastian Pokutta}\hfill{\ttfamily pokutta@zib.de}\\
{\small\emph{Berlin Institute of Technology and Zuse Institute Berlin}}

\vspace{3mm}

\begin{abstract}
  Fast domain propagation of linear constraints has become a crucial
  component of today's best algorithms and solvers for mixed integer
  programming and pseudo-boolean optimization to achieve peak solving
  performance. Irregularities in the form of dynamic algorithmic
  behaviour, dependency structures, and sparsity patterns in the input
  data make efficient implementations of domain propagation on GPUs
  and, more generally, on parallel architectures challenging. This is
  one of the main reasons why domain propagation in state-of-the-art
  solvers is single thread only. In this paper, we present a new
  algorithm for domain propagation which (a) avoids these problems and
  allows for an efficient implementation on GPUs, and is (b) capable
  of running propagation rounds entirely on the GPU, without any need
  for synchronization or communication with the CPU. We present
  extensive computational results which demonstrate the effectiveness
  of our approach and show that ample speedups are possible on
  practically relevant problems: on state-of-the-art GPUs, our
  geometric mean speed-up for reasonably-large instances is around 10x
  to 20x and can be as high as 180x on favorably-large instances.
\end{abstract}

\section{Introduction}

Given a matrix $A \in \mathbb{R}^{m \times n}$, vectors
$b \in \mathbb{R}^m$, $c \in \mathbb{R}^n$, and a subset
$I \subseteq N = \{1, \ldots , n\}$, a \emph{mixed integer linear
  program} (MIP) is an optimization problem that can be written as
\begin{equation}
\label{eq:MIP}
\min \{c^Tx \; | \; Ax \leq b, x \in \mathbb{R}^n , x_j \in \mathbb{Z} \; \text{for all} \; j \in I\}.
\end{equation}

Typically, MIPs are $\mathcal{NP}$-hard to solve, but surprisingly
fast algorithms exist in
practice~\cite{AchterbergWunderling2013,KochMartinPfetsch2013}. The
most successful methods to solve MIPs are the \emph{branch-and-bound}
algorithm and its extensions. The main idea of this algorithm is to
split the original problem into subproblems which are easier to solve
(\emph{branching}). By repeatedly branching on the subproblems, a
search tree is obtained. At each subproblem, the \emph{bounding} step
uses relaxations to compute lower bounds and prune suboptimal nodes of
the tree in order to avoid enumerating exponentially many
subproblems. Relaxations are typically obtained by dropping
integrality requirements on variables $x_j$ for $j \in I$, at which point
a \emph{linear program} (LP) is obtained and solved by, e.g., the
simplex algorithm \cite{NemhauserWolsey1988}.

One of the supplementary techniques used to improve the initial
problem formulation and decrease the size of the branching tree is to
limit the domains of the variables $x_j$ to the value assignments
which can be completed to \emph{feasible solutions} of a given
(sub)problem. The process of computing these ranges is called
\emph{domain
  propagation}~\cite{BelottiCafieriLeeLiberti2010,Achterberg2009}.

Whereas many areas of applied mathematics (e.g., Deep Learning) have
strongly profited from the increased computing power of
parallel coprocessors like \emph{Graphics Processing Units}
(GPUs), success in using these resources to solve MIPs has been very
limited~\cite{BoyerEtal2017}.
The main challenges for developing GPU-accelerated MIP techniques have
been (a) highly sparse input data with irregular sparsity patterns;
(b) inherent sequential properties of many of the best methods used in
the field; and (c) their non-uniform algorithmic behavior. On top of
this, many of the existing algorithms were designed with a sequential
execution model in mind, which does not allow to transform them easily
to an efficient GPU implementation.

In this paper, we focus on domain propagation and propose an
efficient algorithm for GPUs.  In addition to being
able to obtain results faster with this algorithm under almost all
circumstances, we hope that this work can pave the way for the
implementation of other methods or motivate the development of new
optimization methods on GPUs.  Our algorithm is capable of running
propagation rounds entirely on the GPU, without the involvement
of the CPU.  Consequently, this also presents an opportunity to
investigate the use of domain propagation in conjunction with existing
or future methods that also run completely on the GPU. This is
especially interesting given the fact that state-of-the-art MIP
solvers largely do not exploit GPUs during the solving process,
leaving them idle. 

We begin by formally introducing the necessary background and notation.

\subsection{Domain Propagation}
\label{sub:dompropintro}
Domain propagation is an iterative technique used to tighten the bounds of variables at each node of the branch-and-bound tree. It is also used as a pre-processing technique to improve the formulation of a given MIP model~\cite{Savelsbergh1994}.
In the following, we consider linear constraints of the form
\begin{equation}
\label{eq:1}
\underline{\beta} \le a^Tx \le \overline{\beta},
\end{equation}
where $\underline{\beta}, \overline{\beta} \in \mathbb{R} \cup \{-\infty, \infty\}$ are left and right hand sides, respectively, and $a \in \mathbb{R}^n$ is the vector of constraint coefficients.
We assume that variables $x_j$ have lower and upper bounds $\ell_j, u_j \in \mathbb{R} \cup \{-\infty, \infty\}$.
Before stating the formulas for updating variable bounds, we need the following definition.

\begin{definition}[activity bounds]
\label{actsdefinition}
Given a constraint of form~\eqref{eq:1} and bounds $\ell \le x \le u$, we call
\begin{subequations}
\begin{align}
\label{eq:minactivities}
& \underline{\alpha} := \sum_{i=0}^n a_ib_i \text{ with } b_i = 
  \begin{cases}
      \ell_i & \text{if}\ a_i > 0 \\
      u_i & \text{if}\ a_i \le 0
    \end{cases} 
\intertext{the \emph{minimum activity}, and}
\label{eq:maxactivities}
 & \overline{\alpha} := \sum_{i=0}^n a_ib_i \text{ with } b_i = 
  \begin{cases}
      u_i & \text{if}\ a_i > 0 \\
      \ell_i & \text{if}\ a_i \le 0
    \end{cases}
\end{align}
\end{subequations}
the \emph{maximum activity} of the constraint. 
\end{definition}

Then domain propagation is based on three
observations~\cite[Sec.~7.1]{Achterberg2009} and translate into the
following algorithmic steps.
\begin{enumerate}[1:]
\item\label{dp:stepi} If $\underline{\beta} \le \underline{\alpha}$ and $\overline{\alpha} \le \overline{\beta}$, then the constraint is redundant and can be removed.
\item\label{dp:stepii} If $\underline{\alpha}>\overline{\beta}$ or $\underline{\beta}>\overline{\alpha}$, then the constraint cannot be satisfied and hence the entire (sub)problem is infeasible.
\item\label{dp:stepiii} Let $x$ satisfy \eqref{eq:1}, then for all $j=1,\ldots,n$ with $a_j > 0$,
\begin{subequations}
\begin{align}
\label{eq:bound_candidates_pos}
& \frac{\underline{\beta} - \overline{\alpha}}{a_j} + u_j \le x_j \le \frac{\overline{\beta} - \underline{\alpha}}{a_j} + \ell_j,
\intertext{and for all $j=1,\ldots,n$ with $a_j < 0$,}
\label{eq:bound_candidates_neg}
& \frac{\overline{\beta} - \underline{\alpha}}{a_j} + u_j \le x_j \le \frac{\underline{\beta} - \overline{\alpha}}{a_j} + \ell_j.
\end{align}
\end{subequations}
If $x_j$ is integral, lower bounds can be rounded up and upper bounds can be rounded down to strengthen them further.
\end{enumerate}

If Steps~\ref{dp:stepi} and~\ref{dp:stepii} are not applicable, the propagation algorithm
computes new bounds $\lnew_j, \unew_j$ in Step~\ref{dp:stepiii}. If
$\lnew_j > \ell_j$ or $\unew_j < u_j$ for some $j$, then the bounds of
variable $x_j$ are updated.
To apply domain propagation to a MIP formulation, one simply applies these steps to all of its constraints.

An actual implementation may skip Steps~\ref{dp:stepi} and~\ref{dp:stepii} without changing the result.
This is because for redundant constraints Step~\ref{dp:stepiii} correctly detects no bound
tightenings, and for infeasible constraints, Step~\ref{dp:stepiii} leads to at least one
variable with an empty domain, i.e., $\lnew_j > \unew_j$.

Let us consider what happens when the steps are applied to all
constraints of the system and some bound changes have been
found. First, note that the formulas for computing new variable bounds
in \eqref{eq:bound_candidates_pos} and \eqref{eq:bound_candidates_neg}
depend on the activity bounds $\underline{\alpha}$ and
$\overline{\alpha}$. The activity bounds themselves depend on $\ell_j$
and $u_j$. Second, note that several constraints typically share the
same variable $x_j$. In this case, updating the bounds of $x_j$ during
the processing of one constraint changes the bound and activity values
for all other constraints that contain $x_j$. Propagating the affected
constraints again can then lead to further tightenings.

Hence, domain propagation is usually applied iteratively. Each
iteration of applying Steps~\ref{dp:stepi} to~\ref{dp:stepiii} to all or a subset of the
constraints is also called a \emph{propagation round}. If no bound
updates are found during a given round, then no further improvements
are possible, and the algorithm terminates.

As pointed out in \cite{BelottiCafieriLeeLiberti2010}, iterated domain propagation can be interpreted as a fixed-point iteration in the space of variable and activity bounds, and there exists a unique limit point of this fixed-point iteration.
Iterated domain propagation converges to this well-defined result, however, not necessarily in finite time~\cite{BelottiCafieriLeeLiberti2010}.
Also, convergence can be very slow in practice~\cite{Achterberg2009}.
The most common way to deal with non-finite or slow convergence in practice is to terminate when the improvements made fall below a specified threshold.
Such tolerance-based termination criteria yield finite termination, but may fail to compute the tightest bounds possible.

\subsection{Graphics Processing Units}

In this section we briefly introduce GPU concepts used throughout the paper. Although the key concepts are independent, we restrict our notation to NVIDIA's GPUs and the CUDA programming model. GPUs are massively parallel processors capable of running hundreds of thousands of threads. Threads are grouped into \emph{warps} in hardware (usually 32 threads each) and forced to execute in a SIMD (Single Instruction Multiple Data) fashion and share the same resources. On the programming level, threads are divided into \emph{thread blocks}. Different thread blocks are scheduled for execution independently, but they all have access to the same \emph{global memory}. Threads inside a single thread block can be synchronized and a given amount of exclusive \emph{shared memory} is available to each block. Accesses to shared memory are usually much faster than accesses to global memory.

%\subsection{CSR Sparse Matrix Storage Scheme}

%To store a sparse matrix $A$, we adopt a ubiquitously used \emph{Compressed Sparse Row (CSR)} %storage scheme. This scheme consists of three arrays: $values$, $cols$ and $row\_delimiters$. %Non-zeros of $A$ are stored row-wise in the $values$ array. The column index for each non-zero %is stored in $cols$ array. Finally, the $row\_delimiters$ array stores the indices of beginning %elements for each row.

\subsection{Related Work}

While our motivation to study domain propagation are MIPs and we
restrict ourselves to linear constraints, the concept is more general
and has been rediscovered several times since the 1970s in different
communities. First, it was used in the AI (Artificial Intelligence)
and CP (Constraint Programming) communities, where its origins can be
traced back to the Waltz algorithm~\cite{Waltz1975}.
% In these communities, it is also referred to as \emph{constraint propagation} and \emph{bounds propagation}.
Several other techniques related to domain propagation are widely used in AI and CP and known under the names \emph{domain filtering}, \emph{domain reduction}, \emph{bound reduction}, \emph{range reduction} and \emph{constraint propagation}.
In MIP, domain propagation was first discussed in the context of presolving \cite{Savelsbergh1994,AndersenAndersen1995}. In global optimization and mixed integer non-linear programming, it is most commonly known as \emph{feasibility-based bounds tightening} (FBBT) and was first used in \cite{ShectmanSahinidis1998}.
% In the Global Optimization community, domain propagation was discussed in \cite{go_1} and  improved in \cite{go_2}.

The literature on previous efforts to successfully exploit GPUs in
exact MIP methods is scarce.  We refer to the recent survey by Boyer
et al.~\cite{BoyerEtal2017} and references therein and we only provide a very brief
summary here.  Most work in this direction has focused on the simplex
method (especially its linear algebra), on dynamic programming
approaches for solving the knapsack problem, and on branch-and-bound
implementations for a few special problem classes. Somewhat more
attention was paid to GPU implementations of metaheuristic methods
which aid exact methods, as their compute tasks seem more amenable to
an efficient implementation on GPUs.
% For more details, the interested reader is referred to a recent survey on the application of GPUs to linear and mixed-integer programming by Boyer et al. in \cite{BoyerEtal2017}.
 
\subsection{Contribution}

To the best of our knowledge, this paper presents the first algorithm
for domain propagation on GPUs. We discuss and address two
main sources of irregularity that challenge efficient GPU
implementations:
\begin{inparaenum}
  \item the dynamic behaviour of the existing CPU-based
algorithms, and
\item the highly sparse and irregular structure of the
constraint matrix~$A$.
\end{inparaenum}
We propose a new algorithm which is better suited to the
\emph{throughput-based} model of GPUs.

To deal with the highly
irregular structure of the input matrix $A$, we found inspiration in
existing work on parallelizing sparse linear algebra and demonstrate
that some of these ideas can be carried over to domain propagation.
As a result, we obtain an efficient method that runs entirely on the
GPU and yields significant speedups for domain propagation on the
MIPLIB~2017 test bed, the \emph{de-facto} standard for MIP solvers. The computational results are presented for both double- and single-precision arithmetic. Finally, we validate our baseline algorithms against the domain propagation implementation in the CPU-based PaPILO presolve library \cite{GamrathEtal2020OO}.

The rest of the paper is organized as follows. In Section~\ref{sec:parallelization} we discuss existing state-of-the-art implementations of domain propagation and the challenges they present for efficient GPU parallelization, alongside the main approach we take in overcoming these challenges. In Section~\ref{sec:impl} we focus on the handling of the irregular (sparse) structure of the constraint matrix and other implementational issues. Finally, in Section~\ref{sec:compres} we present computational results.

\section{Parallelizing Domain Propagation}
\label{sec:parallelization}

We strive to exploit two sources of parallelism from the domain
propagation algorithm: First, we exploit the fact that we can apply
the propagation reductions to each constraint, i.e., to each row of
$A$, independently and thus do so fully in parallel. The repercussions of
this on the algorithm are discussed in Section~\ref{sub:reduced_eff}.
Second, we exploit parallelism inside the processing of each constraint. This itself comes from two sources: (a) we parallelize the computation of minimum and maximum activities, see Sections \ref{sub:actscompandspmvs} through \ref{sub:infcontr}, and (b) we parallelize the updating of bounds, see Section \ref{sub:bdcands}.

\subsection{Features of Sequential Domain Propagation}

We first present the main features of the sequential domain
propagation algorithm and discuss why this algorithm is not suited for
efficient implementation on GPUs. CPU-based, sequential
implementations of the domain propagation algorithm follow the
\emph{latency-based} sequential programming model. The main steps of
this algorithm are summarized in Algorithm~\ref{alg:seq_dom_prop}.

\begin{algorithm}
\caption{Sequential domain propagation}
\label{alg:seq_dom_prop}
\begin{algorithmic}[1]
\REQUIRE System of linear constraints $\underline{\beta} \le Ax \le \overline{\beta}$, $\ell \le x \le u$ %, $A \in \mathbb{R}^{m \times n}$, $\underline{\beta},\overline{\beta} \in \mathbb{R}^m$, $\ell,u \in \mathbb{R}^n$
\ENSURE Tightened variable bounds $\lnew \le x \le \unew$
\STATE mark all constraints $c$ in $\underline{\beta} \le Ax \le \overline{\beta}$ \label{markallcons}
\STATE \code{bound\_change\_found} $\gets$ \code{true}
\WHILE{\code{bound\_change\_found}}
\STATE \code{bound\_change\_found} $\gets$ \code{false}
\FOR{\textbf{each} constraint $c$ in $\underline{\beta} \le Ax \le \overline{\beta}$} \label{consloop}
\IF{$c$ marked} \label{ifconsmarked}
\STATE unmark $c$ \label{unmarkcons}
\STATE compute $\underline{\alpha}$, ${\overline{\alpha}}$ via \eqref{eq:minactivities} and \eqref{eq:maxactivities} \label{seq_prop:compacts}
\IF{can $c$ propagate} \label{canconspropagate}
\FOR{\textbf{each} variable $v$ in $c$}{}
\IF{can $v$ be tightened} \label{canvartighten}

\STATE compute $\lnew_v$, $\unew_v$ via \eqref{eq:bound_candidates_pos} and \eqref{eq:bound_candidates_neg}

\IF{$\lnew_v > \ell_v$}
\STATE $\ell_v \gets \lnew_v$
\STATE \code{bound\_change\_found} $\gets$ \code{true}
\ENDIF

\IF{$u_v^{\text{new}} < u_v$}
\STATE $u_v \gets u_v^{\text{new}}$
\STATE \code{bound\_change\_found} $\gets$ \code{true}
\ENDIF

\IF{\code{bound\_change\_found}}
\STATE mark all constraints $c$ with $v$ in $c$ \label{markconswithvar}
\ENDIF

\ENDIF
\ENDFOR
\ENDIF
\ENDIF
\ENDFOR
\ENDWHILE

\RETURN $\lnew$, $\unew$
\end{algorithmic}
\end{algorithm}

Informally speaking, this algorithm starts with the first non-zero entry of $A$, computes everything it can for the corresponding variable, tightens its bounds if possible, and then moves on to the next non-zero. It can stop processing a constraint or a variable early if the sufficient conditions in Line~\ref{canconspropagate} and Line~\ref{canvartighten} are met. This avoids unnecessary work as these checks can be performed before the new bound candidates are computed and compared to the old bounds. In addition, this algorithm takes advantage of the fact that a constraint can trigger propagation of other constraints only if they share at least one variable by implementing a marking mechanism (Lines \ref{markallcons}, \ref{ifconsmarked}, \ref{unmarkcons} and \ref{markconswithvar}).

This irregular behaviour poses challenges for an efficient implementation on a GPU. Threads assigned to units of work that are stopped early would remain idle during the rest of the execution in a given propagation round, leaving hardware underutilized. The remaining threads would potentially be accessing memory far apart, resulting in uncoalesced accesses. Notice that we do not know a priori which parts would terminate early and at which level, or which constraints will be marked in the next round. This induces highly dynamic behaviour both inside a given propagation round and between different rounds as well.

Additionally, notice that the amount of work a given thread would
perform heavily depends on where, if at all, it will terminate
early. Computing $\underline{\alpha}$ and
$\overline{\alpha}$ in Line~\ref{seq_prop:compacts}, e.g., also requires a
loop over the variables in the constraint, see
\eqref{eq:minactivities} and \eqref{eq:maxactivities}. The marking
operation in Line~\ref{markconswithvar} involves iterating over the
columns of $A$ and finding all constraints that contain the variable
in question. The dynamic behaviour of the algorithm described in the
previous paragraph would also make load balancing the work between
different threads difficult.

\subsection{The Price of Parallelism}
\label{sub:reduced_eff}
As mentioned earlier, the constraints of a given system can be propagated independently. Exploiting this parallelism, however, comes at a cost; it will result in a less efficient algorithm in the worst-case scenario. The main reason for this is that a bound change found during the sequential execution of the algorithm becomes immediately available to the propagation of the subsequent constraints. If this bound change triggers propagation of one of the subsequent constraints, this propagation can be done during the same round. In the parallel case however, all constraints are propagated independently, hence propagation triggered in this way would have to wait until the next propagation round. 

The worst case of such a propagation pattern is a \emph{cascading propagation pattern}, where Constraint 1 triggers propagation of Constraint 2, then Constraint 2 triggers Constraint 3, and so forth. A sequential implementation could propagate this pattern in one round, while a parallel implementation that propagates each constraint independently would require $m$ rounds, where $m$ is the number of constraints in the system. In the best-case scenario, the parallel and sequential algorithm propagate the system in the same number of rounds.
% Note that the necessary condition for a constraint to be able to trigger propagation of another constraint is that they share at least one variable. 

To roughly gauge this ``price of parallelism'' effect, we conducted a preliminary
experiment over \num{893}~instances from the MIPLIB~2017 test
set~\cite{GleixnerEtal2019} on which both our parallel and sequential
implementations converge to identical results.
In this experiment, the average number of propagation rounds of the sequential
implementation was \num{3.1}, but increased to \num{4.4}~rounds on average for the parallel
implementation, hence an average increase by a factor \num{1.4}.
However, the maximum increase observed for an instance was as large as
\num{22.0}.

\subsection{A GPU-Targeted Parallelization}
\label{sub:gpuparallelalg}
In this section, we describe the proposed GPU-parallel domain propagation algorithm and discuss why it better matches the GPU model. Finally, we highlight how the missing features of this algorithm, compared to the sequential algorithm, are mostly remedied by the massively parallel GPU model.

The parallel algorithm we propose for implementation on the GPUs is
shown in Algorithm~\ref{alg:gpu_dom_prop}. It does not contain the
early termination checks and the marking mechanism of the Algorithm~\ref{alg:seq_dom_prop}. This
means that at the expense of performing more computations we avoid the
irregular behaviour of the CPU version induced by these checks. Taking
constraint marking as an example, note that unmarked constraints would
not be processed by the CPU algorithm, while the GPU algorithm would
process such constraint even though it cannot yield improved
bounds. What we gain though is an algorithm with a static sequence of
computations that can exploit the massively parallel GPU model.

\begin{algorithm}
\caption{GPU-parallel domain propagation (schematic)}
\label{alg:gpu_dom_prop}
\begin{algorithmic}[1]

\REQUIRE System of linear constraints $\underline{\beta} \le Ax \le \overline{\beta}$, $\ell \le x \le u$ %, $A \in \mathbb{R}^{m \times n}$, $\underline{\beta},\overline{\beta} \in \mathbb{R}^m$, $\ell,u \in \mathbb{R}^n$
\ENSURE Tightened variable bounds $\lnew \le x \le \unew$
\STATE \code{bound\_change\_found} $\gets$ \code{true}
\WHILE{\code{bound\_change\_found}} \label{gpu_prop:rounds}
\FOR{constraint $c$ in $A$ \textbf{parallel}}
\STATE compute $\underline{\alpha}$, ${\overline{\alpha}}$ via \eqref{eq:minactivities} and \eqref{eq:maxactivities} \label{gpu_prop:compacts}
\ENDFOR

\FOR{constraint $c$ in $A$ \textbf{parallel}}
\FOR{variable $v$ in $c$ \textbf{parallel}}
\STATE compute $\lnew_v$, $\unew_v$ via \eqref{eq:bound_candidates_pos} and \eqref{eq:bound_candidates_neg} \label{gpu_prop:compbdcands}
\IF{$\lnew_v > \ell_v$}
\STATE $\ell_v \gets \lnew_v$
\STATE \code{bound\_change\_found} $\gets$ \code{true}
\ENDIF

\IF{$u_v^{new} < u_v$}
\STATE $u_v \gets u_v^{\text{new}}$
\STATE \code{bound\_change\_found} $\gets$ \code{true}
\ENDIF
\ENDFOR
\ENDFOR

\ENDWHILE

\RETURN $\lnew$, $\unew$
\end{algorithmic}
\end{algorithm}

Notice the shift in the way the two algorithms progress through the computation. The CPU-based Algorithm~\ref{alg:seq_dom_prop} starts with the first constraint and its first variable and continues processing that variable until no more processing is possible. It then moves to the next variable and the process is repeated. Hence, the activities $\underline{\alpha}$ and $\overline{\alpha}$ are computed for the first constraint, then all the variables in that constraint are processed, then the process is repeated on the remaining constraints. The GPU-based Algorithm~\ref{alg:gpu_dom_prop} instead starts with computing $\underline{\alpha}$ and $\overline{\alpha}$ for \emph{all} constraints. Then it computes new bound candidates for \emph{all} non-zeros of $A$. A useful analogy to graph algorithms here is that the CPU Algorithm \ref{alg:seq_dom_prop} resembles a \enquote{depth-first search} and the GPU algorithm resembles a \enquote{breadth-first search} of a graph. In terms of the GPU model, the new algorithm allows for better load balancing, maximizing the \emph{throughput}.

This parallel algorithm also allows us to access large parts of necessary memory in a coalesced way. However, to understand how the memory is accessed, we also need to take in account the sparsity pattern and the storage scheme of $A$. Thus, we discuss memory accesses in Section~\ref{sec:impl} which deals with the irregular structure of $A$.

For an example of why our proposed algorithm leads to better load
balancing in the parallel case, notice that all variables in a given
constraint share the same $\underline{\alpha}$ and
$\overline{\alpha}$. If we first use all available threads to
precompute the activities, then synchronize and perform bound updates,
no threads are left idle and no computation is repeated. Otherwise,
either all the threads compute the same activity values (best-case
scenario we get the result in the time a single thread takes to
compute the activity) or some of them are idle and waiting for others
to compute the activities before performing the bound
updates. Additionally, having a group of threads cooperate on
computation of activities will lead to a faster algorithm. This is
because the computation of activities allows for some parallelism to
be exploited; see Sections \ref{sub:actscompandspmvs} through \ref{sub:infcontr}. 

We now highlight how the \emph{throughput-based} GPU programming model partly
remedies the lack of early termination checks of Algorithm
\ref{alg:seq_dom_prop}. Let us only consider the constraint marking
feature; the conclusion drawn will be analogous for other early
termination checks.

Assume we have $m$ constraints in the system, $k$ of which are marked
for propagation, where $k \le m$. Furthermore, assume that the costs
of propagating one constraint on the hardware where the sequential and
parallel algorithms are run are $C_{\text{seq}}$ and
$C_{\text{par}}$, respectively, and that the parallel algorithm has
$p$ processing units for parallel computation at its disposal. The
sequential algorithm would only propagate the $k$ marked constraints
and thus finish the propagation with the cost of
$k \cdot C_{\text{seq}}$. The parallel algorithm, on the other hand,
would propagate the system with the cost of
$\lceil\frac{m}{p}\rceil \cdot C_{\text{par}}$. For the case of
$m \le p$, it would not matter for the parallel algorithm whether it
propagates all the $m$ constraints or only the $k$ marked
constraints. For $m > p$, we pay the price for propagating one
constraint, i.e., $C_{\text{par}}$ for each additional $p$ constraints
that are propagated.

In practice, it is hard to exactly quantify $p$ for GPUs due to the
complexity of their hardware and execution model. However, we can get
a rough idea of what orders of magnitude are involved in the
computation.
%% Our analysis of \num{1008} instances
%% from the MIPLIB~2017 test set shows that the average numbers for the
The \num{987} instances from our MIPLIB~2017 test bed from Section~\ref{sub:testset} show on average
\num{118514} constraints, \num{64611}
variables, and \num{1226730} non-zeros, respectively. On the other
hand, modern GPUs offer order of tens of thousands of threads running
in parallel. Thus, given the size of practically relevant problems and
the amount of parallelism modern GPUs offer, we can expect that the
price for extra amounts of work in the parallel algorithm will be
mostly negligible.

Further properties of GPUs help in reducing the cost we pay for extra
computations. For example, the GPU hardware is optimized to combine
multiple memory accesses into a single transaction (\emph{memory
  coalescing}). As will be discussed in Section~\ref{sec:impl}, the
GPU algorithm allows us to access large parts of necessary memory in a
fully coalesced manner. So the memory to be operated on by the extra
threads will often be already loaded. In low arithmetic intensity
algorithms like domain propagation memory accesses are typically
dominant parts of GPU implementations in terms of runtime. Hence, we
do not expect these extra arithmetic operations to have a significant
effect on the total runtime of the algorithm; our computational
results confirm this.

\section{Handling Irregular Structure of the Constraint Matrix}
\label{sec:impl}

In practice, constraint matrices $A$  coming from MIPs are very
sparse. To store $A$, we adopt a ubiquitously used \emph{Compressed Sparse Row (CSR)} storage scheme \cite{GreathouseDaga2014}. Additionally, MIPs often contain so-called \emph{connecting
  constraints} which are very dense. Consequently, even though matrix
$A$ might be very sparse overall, it may contain a few very dense
rows. This poses a challenge for load balancing on GPUs. Consider two
naive approaches of assigning a single thread per constraint and
assigning a warp or block of threads per constraint. In the former
case, threads assigned to dense rows would have much more work to do,
leaving other threads idle. In the latter case, warps or blocks
assigned to rows with a small number of non-zeros would remain
underutilized.

We already discussed how GPU hardware is optimized around coalescing
memory accesses of threads in a warp into one or as few as possible
memory accesses. The approach of assigning one thread per constraint
would perform poorly in this regard as neighboring threads running in
parallel would access memory locations which are not adjacent in the
row-major CSR format. Assigning a warp of threads per constraint, on
the other hand, results in fully coalesced memory accesses: each
thread of a warp takes one non-zero element, which are stored next to
each other in the CSR format.

In this section, we address above-mentioned challenges and present an algorithm which combines good load balancing with coalesced memory accesses.  To introduce the main idea, we first focus on the computation of the minimum and maximum activities $\underline{\alpha}$ and $\overline{\alpha}$ in Section~\ref{sub:actscompandspmvs}, \ref{csradaptive}, \ref{sub:csradaptiveforactivities}, and \ref{sub:infcontr}. The second part of the algorithm, which computes new bound candidates is discussed in Section~\ref{sub:bdcands}.   

\subsection{Similarity Between SpMV and Computation of Activities}
\label{sub:actscompandspmvs}

The first step of the parallel algorithm computes minimum and maximum activities $\underline{\alpha}$ and ${\overline{\alpha}}$ for all constraints of the system via \eqref{eq:minactivities} and \eqref{eq:maxactivities}.
Our main approach for implementing these formulas
is the observation that their efficient implementation for a matrix $A$ on GPUs is
conditioned by the same factors that condition implementations of a
sparse matrix-vector product (SpMV) of $A$ and some right-hand side
vector. SpMVs and their efficient implementations on GPUs have been
extensively studied over the years as they play a crucial role in
computational science. We will then be able to carry over ideas to
overcome challenges described at the beginning of this section for computing
activities.

To support our observation, we rely on the fact that SpMV is bandwidth
bound (i.e., its arithmetic intensity is very low), so good memory
access patterns greatly improve performance
\cite{GreathouseDaga2014}. Notice that arithmetic intensity of matrix-vector products and computing $\underline{\alpha}$ or $\overline{\alpha}$ is the
same. When it comes to memory needed to perform the operations to compute the activities, notice
that each $l_i$ and $u_i$ is accessed exactly once: either for
$\underline{\alpha}$ or for $\overline{\alpha}$. This means that we
need to access each element of $A$, $l$, and $u$ exactly once to implement
\eqref{eq:minactivities} and \eqref{eq:maxactivities}. This is exactly
the same memory needed to compute the matrix-vector product of $A$ with two right hand sides $l$ and $u$.

Greathouse and Daga introduced an algorithm in \cite{GreathouseDaga2014} which addresses issues discussed at the beginning of this section in the context of SpMVs. This algorithm is called \emph{CSR-adaptive}. It handles well both structured and unstructured matrices and explicitly addresses the case of having very long rows in an otherwise sparse matrix. We now briefly introduce the main idea of this algorithm before discussing the changes made to facilitate computation of activities. For a detailed discussion of \textit{CSR-adaptive} see \cite{GreathouseDaga2014}.

\subsection{The CSR-Adaptive Algorithm}
\label{csradaptive}

The main idea of the algorithm is to divide the matrix $A$ into \emph{row blocks}, and have one CUDA threads block work on one row block. If the number of non-zeros in some rows is small, they will be grouped together in one row block. A so-called \emph{CSR-stream} algorithm will then be applied to this row block. If a given row block consists of only one or few rows, a so-called \emph{CSR-vector} variant will be applied.

The CSR-stream algorithm first assigns one thread to each non-zero in the row block and loads them into shared memory. This results in fully coalesced memory accesses. Afterwards, a number of threads is assigned per each row that is present in the row block. These threads then carry out the necessary computations and reductions. The CSR-vector algorithm assigns one warp of threads to the row block which then perform the computations and reductions.

\subsection{Computing Minumum and Maximum Activities}
\label{sub:csradaptiveforactivities}

On the implementation level, we make the following changes to the CSR-adaptive algorithm as presented in \cite{GreathouseDaga2014}. First, instead of working on one right-hand side array, we adapt the algorithm to work on two arrays: $\ell$ and $u$. As part of the same change, the algorithm now outputs two arrays, the minimum and maximum activities.

Second, the arithmetic operations are adjusted to \eqref{eq:minactivities} and \eqref{eq:maxactivities}. Note that this leaves the reductions (i.e., summations) unchanged as it only differs from SpMV in computing the local summands $a_ib_i$.

Third, we allow the CSR-vector algorithm to use all warps in a CUDA
thread block if the rows are extremely long. In this case, each warp
computes partial sums of its elements, after which the partial sums
are reduced in shared memory. In our implementation, we use a
(somewhat arbitrary) 
threshold value of \num{64} to switch between the two variants of
CSR-vector.

Both the CSR-stream and the CSR-vector variant store the computed activities in local shared memory, as they will be utilized in forthcoming computations (see Section~\ref{sub:bdcands}).

\subsection{Numerical Considerations for Computing Activities}
\label{sub:infcontr}

Apart from the standard issues that numerical algorithms based on floating-point arithmetic have to deal with, one special case arises during domain propagation that warrants discussion. Equations \ref{eq:bound_candidates_pos} and \ref{eq:bound_candidates_neg} provide formulas for computing bound candidates of a variable $j$ in a given constraint. Notice that these formulas always explicitly contain the value of $\ell_j$ or $u_j$, even though they could be factored into the activity values:

\begin{subequations}
\begin{align}
\label{eq:minresactivities}
& \underline{\alpha_j} = \sum_{i=0, i \ne j}^n a_ib_i = \underline{\alpha} - a_jb_j \text{ with } b_k = 
  \begin{cases}
      \ell_k & \text{if}\ a_k > 0 \\
      u_k & \text{if}\ a_k \le 0
    \end{cases}, 
\intertext{and}
\label{eq:maxresactivities}
 & \overline{\alpha_j} = \sum_{i=0, i \ne j}^n a_ib_i = \overline{\alpha} - a_jb_j \text{ with } b_k = 
  \begin{cases}
      u_k & \text{if}\ a_k > 0 \\
      \ell_k & \text{if}\ a_k \le 0
    \end{cases}.
\end{align}
\end{subequations}

The values $\underline{\alpha_j}$ and $\overline{\alpha_j}$ are called \emph{residual activities}. Literature on domain propagation which does not focus on implementation usually defines the Equations \ref{eq:bound_candidates_pos} and \ref{eq:bound_candidates_neg} in terms of these values rather than the minimum and maximum activities (see \cite{Achterberg2009} for example).

The main reason for not using residual activities directly in practical implementations of domain propagation is performance. Namely, residual activities need to be computed for each non-zero of the system. On the other hand, minimum and maximum activities are defined per constraint of the system. The artificial step of coming back to the residual activities costs almost nothing during the computation of new bound candidates (see Section \ref{sub:bdcands}), as the additional lower or upper bound value and the coefficient are readily available and already used at this point.

However, the case when infinite bounds are involved must be handled with care.
%% Looking from the point of view of a given non-zero element in the system belonging to the constraint $i$ and variable $j$, what happens in essence is that its contribution is first added to the activities of row $i$, but is then subtracted away during the new bounds computation. This leaves the code susceptible to numerical issues when infinite bounds are present.
Consider a constraint $i$ with a number of variables whose bounds have finite values and one variable $j$ whose bounds are infinite. The maximum and minimum activities of such a constraint are infinite. The residual activities of all but variable $j$ are also infinite. However, the residual activities of varible $j$ will be finite. A difficulty arises when computing the new bound candidates for variable $j$, where an infinite value would first need to be added and then subtracted from the result.
(Notice that the case where constraint $i$ contains more than one variable with infinite bounds is easier to handle, because the minimum and maximum activity and all the residual activities are infinite.)

Existing domain propagation implementations, such as the implementation in the PaPILO presolver \cite{GamrathEtal2020OO}, handle this case by keeping track of the number of infinite contributions to each of the minimum and maximum activities separately from the finite part of the activity sums. Then, the special case of having exactly one infinity contribution can be detected and the correct finite value computed. We take the same approach, but unlike sequential algorithms where keeping track of and aggregating a counter is a trivial problem, we have GPU threads accessing non-zeros of the constraint in parallel.

We observe that the problem of computing the number of infinity contributions in a constraint is a reduction problem equivalent to the computation of the activity value itself. Moreover, the memory that needs to be loaded from GPU global memory to compute these two reductions is exactly the same, we just need to use it in a different way: for activities, we compute the summands $a_ib_i$ from Equations \ref{eq:minactivities} and \ref{eq:maxactivities}, while to compute the number of infinity contributions we check $b_i$, and set the summand value to 1 if $b_i$ is infinite, and to 0 otherwise. These summands are then summed in an equal way. Hence we can compute the number of infinity contributions by extending the reductions we already use. This comes at the expense of additional shared and register memory and some computations, however, no additional expensive global memory loads are needed. As already discussed, the activity computation kernel is highly memory-bound, so this approach works in our favor.

\subsection{Computing New Bound Candidates}
\label{sub:bdcands}

The new bound candidates for each variable of a constraint $\underline{\beta} \le a^Tx \le \overline{\beta}$ are computed by \eqref{eq:bound_candidates_pos} and \eqref{eq:bound_candidates_neg}. This operation \emph{maps} each non-zero element $a_{ij}$ of $A$ to its lower and upper bound candidates for variable $j$. 

Our variant of the \textit{CSR-adaptive} algorithm for computing activities from Section~\ref{sub:csradaptiveforactivities} works on a granularity level of one thread per non-zero element of $A$, which is the same granularity we need to map non-zero elements of $A$ to new bound candidates. Moreover, each thread loads the values $a_j$, $l_j$ and $u_j$ into shared memory, and the computed $\underline{\alpha}$ and $\overline{\alpha}$ are also saved in shared memory. Thus, we extend the activities kernel to also compute \eqref{eq:bound_candidates_pos} and \eqref{eq:bound_candidates_neg} after the computation of activities is complete. In most cases, this implementation benefits from reusing the values that are kept in shared memory and avoids expensive GPU global memory loads.   

Finally, once individual threads have computed the corresponding lower and upper bound candidate, they are compared to the best current bounds and updated if necessary. Because a given variable can be present in several constraints, it is possible for multiple threads to hold bound candidates for the same variable. This can lead to race conditions if such threads attempt to update the same variable's bounds in parallel. To deal with these race conditions, we use CUDA's atomic operations to update bounds.

Having all threads of the GPU perform atomic operations at the same time can incur a significant performance cost. In order to limit the number of necessary atomic operations, we exploit the fact that the bounds are strictly improving during the course of the algorithm. For a variable $j$ with old bounds $\ell_j$ and $u_j$ and new bounds $\ell_j^{\text{new}}$ and $u_j^{\text{new}}$, the following holds: $\lnew_j \ge \ell_j$ and $u_j^{\text{new}} \le u_j$. Each thread has its own new bound candidates $\ell_{i,j}^{\text{cand}}$, $u_{i,j}^{\text{cand}}$ for the constraint $i$ and variable $j$ it is processing, but all threads assigned to this variable have the same bound values $\ell_j$ and $u_j$ from the previous round. However, because $\lnew_j \ge \ell_j$ then the following holds: $\ell_{i,j}^{\text{cand}} \ge \ell_j$, or the bound candidate cannot become the new bound. An equivalent argument can be made for the upper bound. This means that we can check if the candidate improves on the bounds from previous round first, and perform an atomic operation to update the actual bound only if it does. In other words, the algorithm discards useless candidates directly and only uses atomics to choose the best of all improvements.
The pseudocode in Algorithm \ref{alg:kernel_prop} summarizes the final algorithm performing one propagation round.

\begin{algorithm}

\caption{Kernel performing one propagation round in Algorithm~\ref{alg:gpu_dom_prop}. \textit{blockIdx} is the index of the current CUDA block the thread executing the code belongs to.}
\label{alg:kernel_prop}
\begin{algorithmic}[1]

\STATE \code{start\_row} $\gets$ \code{row\_blocks[blockIdx]}
\STATE \code{end\_row} $\gets$ \code{row\_blocks[blockIdx + 1]}
\STATE \code{nnz\_block} $\gets$ \code{row\_ptrs[end\_row] - row\_ptrs[start\_row]}
% if more than one row in block

\IF{\code{end\_row} - \code{start\_row} $>$ \code{1}}
\STATE compute $\underline{\alpha}_{\text{row}}$, ${\overline{\alpha}}_{\text{row}}$ by \textit{CSR-stream}
\ELSE 
\IF{\code{nnz\_block} $<$ \code{length\_threshold}}
\STATE compute $\underline{\alpha}_{\text{row}}$, ${\overline{\alpha}}_{\text{row}}$ by \textit{CSR-vector} with one warp
\ELSE
\STATE compute $\underline{\alpha}_{\text{row}}$, ${\overline{\alpha}}_{\text{row}}$ by \textit{CSR-vector} with all warps
\ENDIF
\ENDIF
\STATE \code{\_\_syncthreads()}
\STATE \textit{// each thread knows the row idx $i$ and column idx $j$ of the non-zeros it is processing}
\STATE compute $\ell_{i,j}^{\text{cand}}$, $u_{i,j}^{\text{cand}}$ via \eqref{eq:bound_candidates_pos} and \eqref{eq:bound_candidates_neg}

\IF{$\ell_{i,j}^{\text{cand}} > \ell_j$}
\STATE $\ell_j \gets \code{atomicMax}(\ell_j, \ell_{i,j}^{\text{cand}})$
\ENDIF

\IF{$u_{i,j}^{\text{cand}} < u_j$}
\STATE $u_j \gets \code{atomicMin}(u_j, u_{i,j}^{\text{cand}})$
\ENDIF

\end{algorithmic}
\end{algorithm}

\subsection{Performance Effects as a Function of the Input}
\label{sub:perf_effects_input}

Let us now consider the effect of the shape and the sparsity pattern of the constraint matrix $A$ on the performance of the developed algorithm. We already discussed how separate constraints can be processed independently (fully parallel) to compute the activities
and the new bound candidates. While computing activities, the algorithm also exploits parallelism inside each constraint. However, the amount of parallelism here is limited due to the summation operation.

Suppose now the number of non-zeros of $A$ is fixed. Then, from the point of view of parallel performance and scalability, this step will favor matrices with (a) more rows and (b) fewer non-zeros per row. This is because different rows can be processed fully in parallel, while the computations inside a row have an inherent sequential part that increases with the non-zeros count. 

By contrast, the discussed necessity to use atomic operations to update bounds from the same column means that matrices with more non-zeros per column will have more threads competing for hardware resources, reducing parallel performance and scalability. Different columns, on the other hand, can be processed independently and fully in parallel with no possibility for hardware resources conflict. Therefore, this step will favor matrices with (c) more columns and (d) fewer non-zeros per column. Observe that (a) and (b) are in opposition to (c) and (d), and that the trade-off between them will define their final effect on performance.

Finally, observe that the above ``static'' analysis of constraint matrix shape and sparsity pattern does not take into account the dynamic behavior of domain propagation, which will also affect performance. Two instances with the same shape, the same number of non-zeros and sparsity pattern might still behave differently. For example, one instance might only require a few bound changes to reach the limit point, while the other one might require thousands in many propagation rounds, possibly changing the part of the algorithm with dominating effect on performance; the bound changes in one instance might all happen to be from the same column, resulting in a bottleneck due to many atomic operations, while the other might need no atomic operations at all because all bound changes happen to be from different columns. Even if the two instances have the same bound changes, the timing of the changes will define the amount of hardware resources conflicts the atomic operations will have, e.g., if they are evenly distributed throughout the propagation rounds this might reduce the amount of conflicts, while if the majority is clustered in the first propagation round this might result in a high amount of conflicts. To conclude, the true potential for parallelism can only be evaluated empirically over a sufficiently heterogeneous test set.

\subsection{The Final Algorithm}
\label{sub:cpugpuloop}

\newcommand{\cpuloop}{\textsl{cpu\_loop}\xspace}
\newcommand{\gpuloop}{\textsl{gpu\_loop}\xspace}
\newcommand{\megaker}{\textsl{megakernel}\xspace}

Algorithm \ref{alg:kernel_prop} performs one propagation round. The execution of rounds is iterated as long as at least one bound change is found. If no bound change is found, the limit point is reached and the algorithm terminates. We provide three implementations of this behaviour.

The first approach launches a kernel with one thread block and one thread to iterate through propagation rounds. This kernel then uses CUDA's \textit{dynamic parallelism} feature to spawn the kernel implementing Algorithm \ref{alg:kernel_prop} for each round, as many times as necessary. The kernels communicate through a boolean variable stored in GPU global memory to check if at least one bound change was found during a particular propagation round. This way, the CPU does not have to communicate with the GPU during the execution at all and can continue doing other computations while the bounds propagation is running on the GPU. We denote this implementation as \gpuloop.

The second approach is to move the loop iterating through propagation rounds to the CPU. This means that minimal communication between the GPU and the CPU is necessary between propagation rounds: a boolean variable holding the information if at least one bound change was found during a given round. With this algorithm, the CPU has to perform a minimal amount of work during the propagation, namely, check one boolean variable in a loop and either invoke another kernel execution or otherwise terminate.  However, this is overall faster than the first variant for the following reason: GPUs are built to take advantage of massive parallelism through concurrent threads, but any of those individual threads is typically orders of magnitude slower than a CPU thread. Small as it is, the sequential point of iterating through the propagation rounds and checking if any bound changes have been found has a noticable effect on performance. We denote this implementation as \cpuloop.

The third approach is to avoid the situation where a kernel with one thread and one block has to be launched by launching only one kernel for the entire duration of the execution. This kernel is launched with a number of blocks and threads that allows full GPU occupancy but still does not need to synchronize with the CPU. On the implementation level, this is achieved by combining the \textit{grid-stride loop} kernel design with the \textit{cooperative groups} feature to achieve grid-wide synchronization within the kernel. However, this still does not eliminate the sequential point of the algorithm, as all threads but one remain idle during the synchronization phase. Even though this implementation did outperform the first version with a single-threaded kernel and dynamic parallelism by a measurable amount for some specific instances, it performed worse overall. We denote this implementation as \megaker.

In Appendix \ref{ap:gpucpuloop} we present results quantifying the difference in performance of the three implementations. In the main experiments of the paper presented in Section \ref{sec:compres}, we use the best-performing \cpuloop variant.

\section{Computational Experiments}
\label{sec:compres}

\newcommand{\cpuseq}{\textsl{cpu\_seq}\xspace}
\newcommand{\cpuomp}{\textsl{cpu\_omp}\xspace}
\newcommand{\gpuatomic}{\textsl{gpu\_atomic}\xspace}

\renewcommand{\tesla}{\textsl{V100}\xspace}
\newcommand{\titan}{\textsl{TITAN}\xspace}
\newcommand{\quadro}{\textsl{P400}\xspace}
\newcommand{\amdtr}{\textsl{amdtr}\xspace}
\newcommand{\xeon}{\textsl{xeon}\xspace}
\newcommand{\super}{\textsl{RTXsuper}\xspace}
\newcommand{\planck}{\textsl{i7-9700K}\xspace}

In this section, we present our experimental setup for studying the behavior of
the proposed algorithms and discuss the results obtained.

\subsection{Test Set}
\label{sub:testset}

To benchmark the algorithms, we use the MIPLIB~2017 collection set with
\num{1065} MIP instances, which is currently the largest and most widely
adopted general testbed for MIP algorithms~\cite{GleixnerEtal2019}. The number of non-zeros in the constraint matrix $A$ of this test set ranges
from~\num{3} to \num{256963402}.
Due to the errors coming from the open-source file reader and other issues, we were not able to obtain results for 78 instances, leaving the
set at \num{987}~instances.

As explained in Section~\ref{sub:dompropintro}, domain propagation
may not converge to its limit point in finite time. We have set a limit on the maximum number of propagation rounds in our algorithm at \num{100}. There are \num{30} instances that did not converge within this limit, and we do not use them in performance comparisons.  Additionally, some instances
may run into numerical difficulties due to floating-point arithmetic and other problems. This happened to \num{64} of the instances in this set. Out of \num{987} instances, this leaves us with \num{893} instances which terminate successfully and converge to the same limit point (see Section \ref{sub:comparison_metric} for the definition of convergence to the same limit point). 

Because small instances do not provide enough workload to justify their
parallelization on GPUs, we further eliminate all \num{107}~instances that have less
than \num{1000}~variables and \num{1000}~constraints. Finally, this leaves us with a set of
\num{786}~instances that are used for performance comparisons.

All the results presented in Section \ref{sec:compres} preserve the original ordering of constraints and variables in the instances. The study which shows to what extent the ordering of constraints and variables influences the performance of our algorithms is presented in Appendix \ref{ap:ordering}. 

\newcommand{\ksubset}[2]{$[#1\text{k},#2\text{k})$}
\newcommand{\setn}[1]{\mbox{\textsl{Set-#1}}}
To better capture the performance of the algorithms with respect to the size of the
input problems, we partition this set into eight subsets of increasing size.
We denote by $[s,t)$ the set of instances that contain less than $t$ variables
and $t$ constraints, but have at least $s$ variables or $s$ constraints.
We consider the partitions \ksubset{1}{10}, \ksubset{10}{20}, \ksubset{20}{40},
\ksubset{40}{80}, \ksubset{80}{160}, \ksubset{160}{320}, \ksubset{320}{640}, and
$[640k,\infty)$, and refer to these sets as \setn{1} to \setn{8}. Number of instances in sets \setn{1} through \setn{8} are: \num{270}, \num{129}, \num{98}, \num{91}, \num{65}, \num{57}, \num{40}, and \num{36}, respectively.

The numbers presented in this section differ slightly for the execution on the \quadro GPU (see Section \ref{sub:hardware}), which has much less resources than other machines we used. This results in insufficient memory and similar problems for a few instances. The differences in numbers are never larger than a single digit.

The numbers for all executions are compiled for runs with double-precision arithmetic. Differences for single-precision executions are discussed separately in Section \ref{sub:floatresults}.

\subsection{Algorithms and Hardware}
\label{sub:hardware}

We compare three algorithms\footnote{Our implementations can be found at: \url{https://github.com/Sofranac-Boro/gpu-domain-propagator}}:
\smallskip
\begin{enumerate}
\item \textbf{\cpuseq} is a sequential implementation of
Algorithm~\ref{alg:seq_dom_prop}. It is always executed on a
single-core, following closely the current state-of-the-art
implementation \cite{GamrathEtal2020OO}. 

\item \textbf{\cpuomp} is a shared memory-parallel implementation of
Algorithm~\ref{alg:seq_dom_prop}. OpenMP is used to parallelize the loop
at Line~\ref{consloop}. To improve load balancing, the set of
constraint indices is pre-processed and only those marked for
propagation are assigned to available threads. OpenMP locks are used
to prevent race conditions while updating bounds. 

\item \textbf{\gpuatomic} is the implementation of Algorithm~\ref{alg:kernel_prop} on GPUs in the variant which iterates through the propagation rounds on the CPU (see Section \ref{sub:cpugpuloop}).
\end{enumerate}
The algorithms are tested on the following architectures:

\begin{itemize}[--]
\item \textbf{\tesla} NVIDIA Tesla V100 PCIe 32GB GPU,
\item \textbf{\titan} NVIDIA Titan RTX 24GB GPU,
\item \textbf{\super} NVIDIA GEFORCE RTX 2080 SUPER 8GB GPU,
\item \textbf{\quadro} NVIDIA Quadro P400 2GB GPU,
\item \textbf{\amdtr} 64-core AMD Ryzen Threadripper 3990X @ 3.30 GHz with 128 GB RAM CPU,
\item \textbf{\xeon} 24-core Intel Xeon Gold 6246 @ 3.30GHz with 384 GB RAM CPU,
\item \textbf{\planck} 8-core Intel i7-9700K @ 3.60GHz with 64 GB RAM
\end{itemize}

The list includes high-end GPUs, such as the data center-grade \tesla and \titan, which is also suitable for desktop computing. The \quadro, on the other hand, is a very low-grade GPU that can often be found in personal use desktops. The \amdtr and \xeon are data center-grade CPU servers, while the \planck machine is a desktop computer. 

The \cpuseq and \cpuomp algorithms are executed in double-precision arithmetic, while the \gpuatomic algorithm is executed in double-precision (discussed in Section \ref{sub:results}) and single-precision (discussed in Section \ref{sub:floatresults}).
The {\cpuomp} algorithm is run on the {\xeon} machine with \num{24}~threads, on the {\amdtr} machine with \num{64}~threads, and on the {\planck} machine with \num{8}~threads.

\subsection{Evaluation Methodology}
\label{sub:comparison_metric}

The main metric we use to compare two executions is \emph{speedup} defined over wall clock time. Average speedups are computed as geometric means.
We choose the {\cpuseq} algorithm running on the {\xeon} machine as a base reference execution. All speedups are computed against this execution, unless stated otherwise.

Both the sequential and GPU versions have a certain amount of initialization work which only needs to be performed once. In the sequential case, e.g., this is computing the column-major CSC storage of the matrix, which is needed for the marking mechanism (see Algorithm~\ref{alg:seq_dom_prop}). For the GPU algorithm, the blocking of the matrix $A$ is precomputed on the CPU (see Section~\ref{csradaptive}), and the necessary memory is sent to the GPU. As in~\cite{GreathouseDaga2014}, we do \emph{not} include these one-time initialization tasks in our time measurement.
Timing starts just before the first propagation round and ends after the last propagation round is executed and the results are available (in CPU memory in the CPU case, in GPU memory in the GPU case).

When comparing the results of two executions, each individual bound is checked, and the results are deemed equal if all bounds are equal within tolerances as follows.  Two values $a$ and $b$ are considered equal if $|a-b| \le (t_{abs}+t_{rel}|b|)$, where $t_{abs}=10^{-8}$ and $t_{rel}=10^{-5}$. The $a$ value is always assigned to the bound of the reference \cpuseq execution, while the $b$ value is always assigned to the exeution being evaluated, e.g., \cpuomp or \gpuatomic.

\begin{figure*}[htbp]
  \centerline{\includegraphics[width=1.0\textwidth]{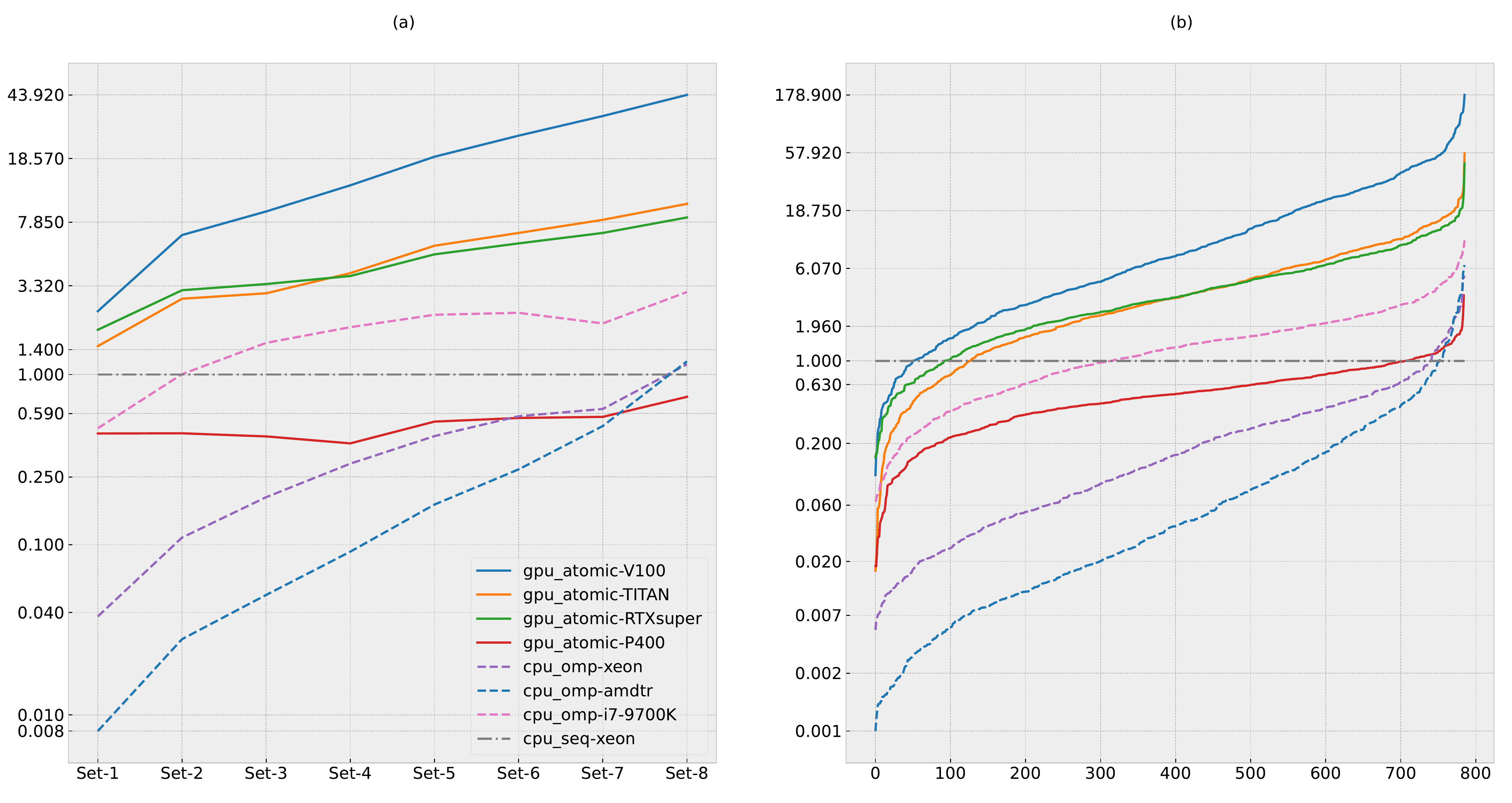}}
  %\vspace{-1.0em}
\caption{The left graph (a) shows geometric means of speedups over the eight
  subsets of instances of increasing size. The right graph (b) shows the
  distribution of speedups by plotting the individual speedup for each instance,
  sorted in ascending order. Each curve belongs to one combination of algorithm
  and machine. All executions are in double-precision floating-point arithmetic. The baseline is the \cpuseq-\xeon execution.}
\label{fig:speedups_double}
\end{figure*}

\subsection{Double-Precision Results}
\label{sub:results}

The speedups that can be observed for the seven algorithm-machine
combinations executed with double-precision floating-point arithmetic are visualized in Figure~\ref{fig:speedups_double}.
Figure~\ref{fig:speedups_double}a) on the left shows the geometric means of speedups for the eight
subsets \setn{1} to \setn{8} of instances with increasing size.
Figure~\ref{fig:speedups_double}b) on the right additionally provides the distributions of the speedups over
all instances sorted in ascending order.
Table~\ref{tab:fullspeedups} provides a summary and reports the average speedup
over all instances and over the subsets \setn{1} to \setn{8}, plus the \num{5}th
percentile, the median, and the \num{95}th percentile speedup.

\begin{table*}
  \caption{Geometric means of speedups across test sets and speedup percentiles for execution with double-precision arithmetic.}
  \label{tab:fullspeedups}
  \centering
\begin{tabular}{@{}lrrrrrrrrrr@{}}
 \toprule
         & \multicolumn{1}{r}{\tesla} & \multicolumn{1}{r}{\titan} & \multicolumn{1}{r}{\super} & \multicolumn{1}{r}{\quadro} & \multicolumn{1}{r}{\amdtr} & \multicolumn{1}{r}{\xeon} & \multicolumn{1}{r}{\planck} \\ 
         & \gpuatomic                 & \gpuatomic                 & \gpuatomic                 & \gpuatomic                  & \cpuomp                    & \cpuomp                   & \cpuomp        \\ \midrule

\multicolumn{6}{c}{\it geometric mean speedups} \\ \midrule

\setn{1} & 2.35                       & 1.47                       & 1.83                       & 0.45                        & 0.01                       & 0.04                      & 0.48            \\ 
\setn{2} & 6.60                       & 2.79                       & 3.13                       & 0.45                        & 0.03                       & 0.11                      & 1.00            \\ 
\setn{3} & 9.08                       & 3.00                       & 3.40                       & 0.43                        & 0.05                       & 0.19                      & 1.53            \\ 
\setn{4} & 12.93                      & 3.94                       & 3.79                       & 0.39                        & 0.09                       & 0.30                      & 1.90            \\ 
\setn{5} & 19.03                      & 5.71                       & 5.08                       & 0.53                        & 0.17                       & 0.43                      & 2.24            \\ 
\setn{6} & 25.31                      & 6.79                       & 5.89                       & 0.56                        & 0.27                       & 0.57                      & 2.30            \\ 
\setn{7} & 33.00                      & 8.11                       & 6.79                       & 0.56                        & 0.50                       & 0.63                      & 2.00            \\ 
\setn{8} & 43.92                      & 10.06                      & 8.37                       & 0.74                        & 1.20                       & 1.15                      & 3.05            \\ \midrule
All      & 7.42                       & 2.98                       & 3.19                       & 0.47                        & 0.04                       & 0.14                      & 1.10            \\ \midrule

\multicolumn{6}{c}{\it percentile speedups} \\ \midrule

5\%      & 0.83                       & 0.37                       & 0.62                       & 0.12                        & 0.003                      & 0.01                      & 0.20            \\ 
50\%     & 7.50                       & 3.35                       & 3.38                       & 0.52                        & 0.04                       & 0.16                      & 1.28            \\ 
95\%     & 51.65                      & 14.80                      & 12.56                      & 1.15                        & 0.80                       & 1.22                      & 3.91           \\ \bottomrule
\end{tabular}
\end{table*}

As can be seen immediately, the \gpuatomic implementation running on the \tesla, \titan, and \super GPUs drastically outperforms both the sequential and the
CPU-parallel executions.
Let us first evaluate the top-performing combination, \gpuatomic on \tesla, in
more detail.
Its average speedup closely follows a linear trend in the size of the instances
from \setn{1} to \setn{8}.
It is always greater than \num{2.35}, and for \setn{8}, which contains all
instances with at least \num{640000} variables or constraints, a speedup
factor of \num{43.92} is achieved.
This shows that, in general, the speedup of the GPU-parallel algorithms
crucially depends on the size of the instances.
Over the whole test set, the average speedup is a factor of \num{7.42}, which
reflects that a significant portion of the test set are small
instances that have limited potential for parallelism.
For 5\% of the instances, however, \gpuatomic on \tesla is at least \num{51.65}~times
faster than the sequential implementation.

By contrast, on the low-end consumer-grade GPU \quadro, the \gpuatomic kernel performs worse than \cpuseq on
all subsets, with an average speedup factor over the whole test set of \num{0.47}. Compared to the other cards, the \quadro has less computational resources available to expoit the parallelism. Additionally, the Volta (\tesla) and Turing (\titan, \super) architectures introduce changes which significantly improve performance \cite{jia2018dissecting} over the Pascal architecture (\quadro) cards.
Nevertheless, keeping in mind that GPUs are currently a resource completely unused by
MIP solvers and that the CPU can do other tasks while the GPU is computing, this result might prove to be interesting even in this very affordable setup, especially as the performance of consumer-grade GPUs will improve over the next years. 

Also the shared memory-parallel algorithm \cpuomp cannot compete with the GPU
algorithms on \tesla, \titan, and \super, which outperform it by a large margin.
The comparatively high cost of managing CPU threads proves to not be justified
by the low arithmetic intensity of the parallel units of work processed by the
threads.
This is especially visible for the \num{24}-thread \xeon and the \num{64}-thread \amdtr executions, which are slower than the \cpuseq on all subsets but \setn{8}. The \num{8}-thread \planck execution, however, outperforms the \cpuseq exeucution on all subsets except for \setn{1}, but its average speedup over the subsets is never larger than \num{3.1}. 

The \gpuatomic executions on \titan and \super GPUs, which both consistently outperform the \cpuseq execution by a large margin, perform quite similarly between themselves. Interestingly, their curves can be seen to cross on both plots. Their similarity in performance is practically relevant given that \super is a cheaper alternative to the high-end \titan GPU, which still achieves comparable performance. 

Figure~\ref{fig:speedups_double}b) allows two further observations.
First, comparing the fastest GPU and CPU executions, the slope of the \tesla execution is noticeably
steeper than the one of \planck, showing that the GPU execution responds better to the growth in parallelism. The same effect can also be seen between the four GPU executions.
Second, the break-even point from which on the \cpuomp on \planck becomes faster than the
sequential implementation lies around the \num{41}st percentile.
By contrast, \gpuatomic on \tesla breaks even at the \num{7}th
percentile.
It wins against the sequential implementation for about \num{93}\% of the
instances.
This highlights that the best GPU-parallel versions are not only significantly
faster than the shared memory-parallel versions, but that they are also much
more robust.

Additionally, for all GPU runs, we can observe a steep increase of the speedup
distribution after the \num{95}th percentile.
This indicates that for particularly favorable structures GPU parallelization can have an
even stronger impact beyond what our general observations over the heterogeneous
MIPLIB~2017 test set suggest.

We also performed the \emph{roofline analysis} \cite{roofline} based on peak bandwidth and performance on the \tesla machine. The heterogeneous MIPLIB~2017 test set contains a number of smaller instances that have a low potential of reaching throughput limits on \tesla, making the roofline analysis unsuitable. Hence, for the purpose of this analysis, we remove all instances with less than \num{250000} non-zeros from the test set. On the remaining \num{349} instances, the average recorded arithmetic intensity is \num{2.96}, with the minimal and maximal recorded values \num{0.26} and \num{17.69}, respectively. The machine balance on the \tesla is \num{8.53}, indicating that the code is, on average, memory-bound. On average, we recorded \num{23.64}\% of attainable performance according to the roofline model, with the minimal and maximal recorded values at \num{1.5}\% and \num{89.14}\%, respectively. These results highlight the highly irregular and dynamic effects that the properties of the input have on the performance of the algorithm (see Section \ref{sub:perf_effects_input}) and reveal both challenges and potential for future improvements.   

\subsection{Single-Precision Results}
\label{sub:floatresults}

For some algorithms running on GPUs, replacing double-precision by single-precision arithmetic may offer significant speedups, if the reduced precision is acceptable from the point of view of correctness of results for the application in question. Additionally, NVIDIA's fast math library (enabled by passing the \texttt{--use\_fast\_math} flag to the \textit{nvcc} compiler) allows for usage of fast hardware implementations of specific operations at the expense of accuracy of results. In this section, we analyze the execution of our \gpuatomic algorithm using single-precision arithmetic both with and without the fast math option. In both cases (as well as for double-precision executions from Section \ref{sub:results}), the \emph{fused-multiply-add} instructions are enabled.

\begin{figure*}[htbp]
  \centerline{\includegraphics[width=1.00\textwidth]{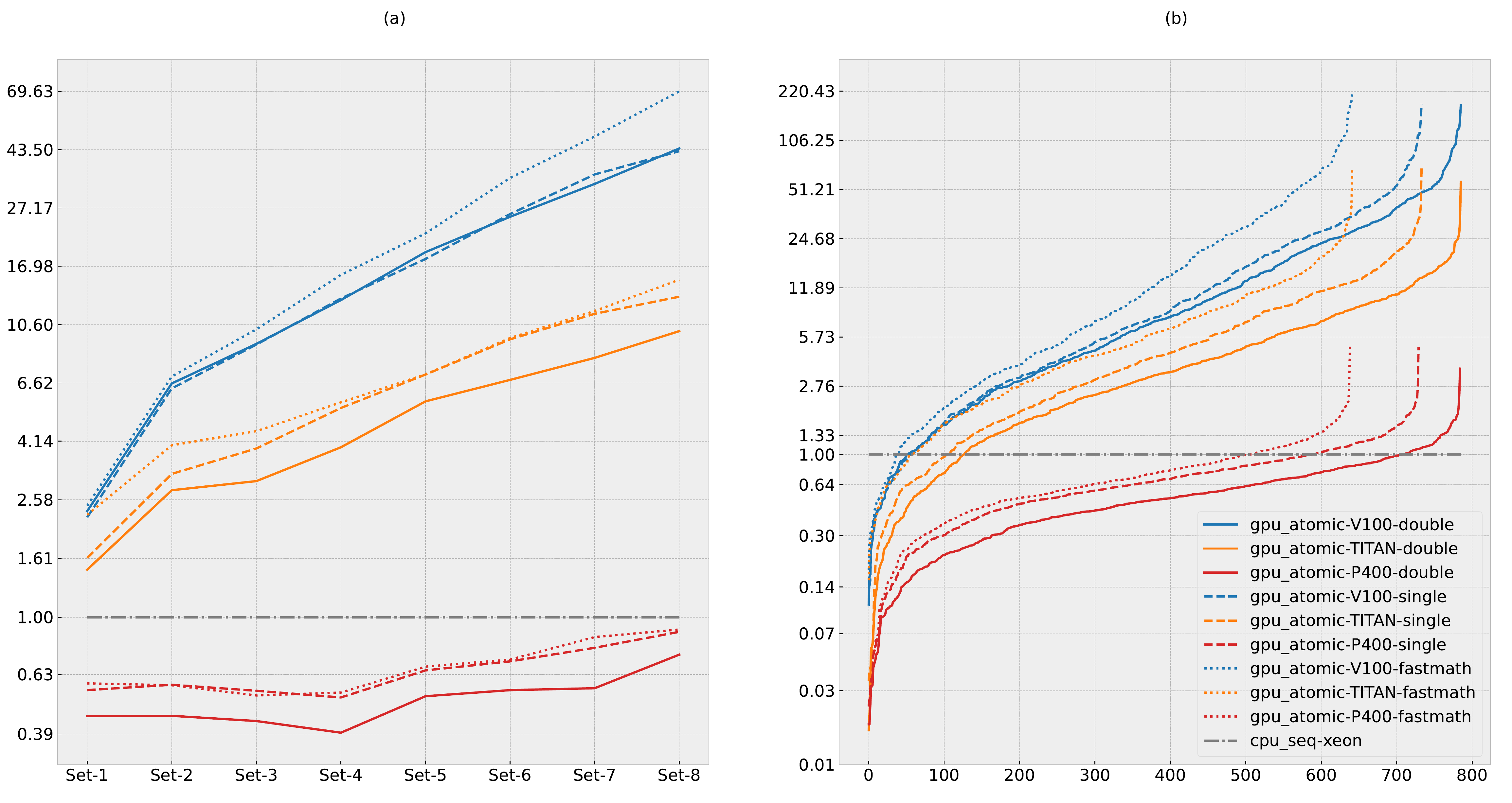}}
  %\vspace{-1.0em}
\caption{The left graph (a) shows geometric means of speedups over the eight
  subsets of instances of increasing size. The right graph (b) shows the
  distribution of speedups by plotting the individual speedup for each instance,
  sorted in ascending order. Double-precision executions are shown in solid lines, standard single-precision executions are shown in dashed lines, while the ``fast math'' mode single-precision executions are shown in dotted lines. The baseline is \cpuseq-\xeon execution in double-precision arithmetic.}
\label{fig:speedups_single}
\end{figure*}

Figure \ref{fig:speedups_single} shows the results for the three GPU machines defined in Section \ref{sub:hardware}: \tesla, \titan, and \quadro. \tesla and \quadro machines both have \num{64} FP32 and \num{32} FP64 cores per streaming multiprocessor \cite{volta,pascal}. \titan, on the other hand, has \num{32} FP32 cores per streaming multiprocessor, with only \num{2} FP64 cores for correctness purposses \cite{turing}. First, we look into the executions without the ``fast math'' option. Out of \num{987} instances, \num{842} converged to the same limit point, \num{27} converged but not to the same limit point, while \num{118} instances hit the maximum number of rounds limit. By contrast, the double-precision executions converge to the same limit point without reaching the maximum number of rounds for \num{893} instances (see Section \ref{sub:testset}).  This shows the effects of reduced precision on the correctness of results. (As before, these numbers are taken from the \tesla machine exeuciton and they might differ by a single-digit number across different machines.)

On the \tesla, we observe that the performance difference between the double- and single-precision executions is minimal: double-precision speedup over the whole test set was \num{7.42}, while the single-precision speedup stands at \num{7.27}. Executions on \titan and \quadro gave a modest speedup: on \titan, the average speedup over the whole set increased from \num{2.98} to \num{3.63}, while on \quadro, it increased from \num{0.47} to \num{0.60}.

The lack of a more significant speedup may be somewhat unexpected, but can be explained by the sparse input data and data structures of the implementations.
Observe that a large part of the memory for the domain propagation kernels is not used for floating-point numbers. Without going into details, we note that the CSR storage format used to represent the sparse constraint matrix contains more integers than floating-point numbers, a memory that is accessed often to implement the indexing logic of the reductions. Due to the low arithmetic intensity of domain propagation, there are usually not many floating-point operations following the indexing.  Additionally, apart from looking at global memory, internal parts of the computations also work on integers. For example, the reductions to keep track of the number of infinity contributions to activities works exclusively on integers, see Section \ref{sub:infcontr}. To summarize, we perform as many reductions on integers as we do on floating-point numbers to compute the activities, plus all the indexing accesses are on integers.

We performed the roofline analysis on single-precision executions on the \tesla machine over the same \num{349} instances used in Section \ref{sub:results}.  The results are similar to the double-precision executions. However, the average recorded percent of peak performance according to the model drops to \num{14.86}\%, with the maximal value recorded at \num{68.9}\%. Average arithmetic intensity was recorded at \num{2.74}, with minimal and maximal values at \num{0.18} and \num{28.85}, respectively. This makes the single-precision code even more memory-bound than the double-precision code. The machine balance value roughly doubles.

Unlike the standard single-precision execution, the combination of ``fast math'' and single-precision execution provided a speedup on the \tesla GPU. Overall, the speedup increased from \num{7.42} to \num{8.54}. On \titan and \quadro, the speedups increased from \num{2.98} to \num{4.38} and from \num{0.47} to \num{0.61}, respectively. For this execution, out of \num{987} instances, \num{736} converged to the same limit point, \num{28} converged but not to the same limit point, while \num{223} instances hit the maximum number of rounds limit. (As before, these numbers are taken from the \tesla machine exeuciton and they might differ by a  single-digit number across different machines.) We can see that the numerics seems to be affected because significantly more instances hit the maximum number of rounds limit, however, the number of instances with inconsistent results only increased by one.

\subsection{Comparison with PaPILO}
\label{sub:papiloperf}

The \cpuseq and \cpuomp algorithms used as the baseline comparison were implemented by the authors for the purposes of this paper. In this section, we evaluate their performance and accuracy against an independent MIP presolver PaPILO, which is distributed as part of the SCIP Optimization Suite 7.0 \cite{GamrathEtal2020OO}. PaPILO provides domain propagation implementation that can run either in single- or multi-thread setups. Our goal is not to provide a detailed and exact computational comparison of the two codes; rather, we want to show that our implementations are of ``reasonable'' performance compared to state-of-the-art domain propagation implementations.

Before interpreting results, we acknowledge that the comparison of the performance of our code against PaPILO is not on equal grounds. First, PaPILO is a much more generic framework, allowing interfacing with different MIP solvers and providing many additional presolving methods, apart from propagation of linear constraints as described in this paper. Even though we disable all other presolving methods but the domain propagation, PaPILO is tuned for the case of multiple methods being applied in combination. Second, PaPILO is meant to be used in conjunction with MIP solvers and will perform reductions useful to them while it performs domain propagation. Specifically, it will remove redundant constraints and variables that are no longer useful to MIP solvers. These operations can not be turned off in PaPILO and will be included in the runtime. Our code, on the other hand, does not try to perform these reductions but focuses solely on domain propagation.

\begin{figure}[htbp]
  \centerline{\includegraphics[width=0.50\linewidth]{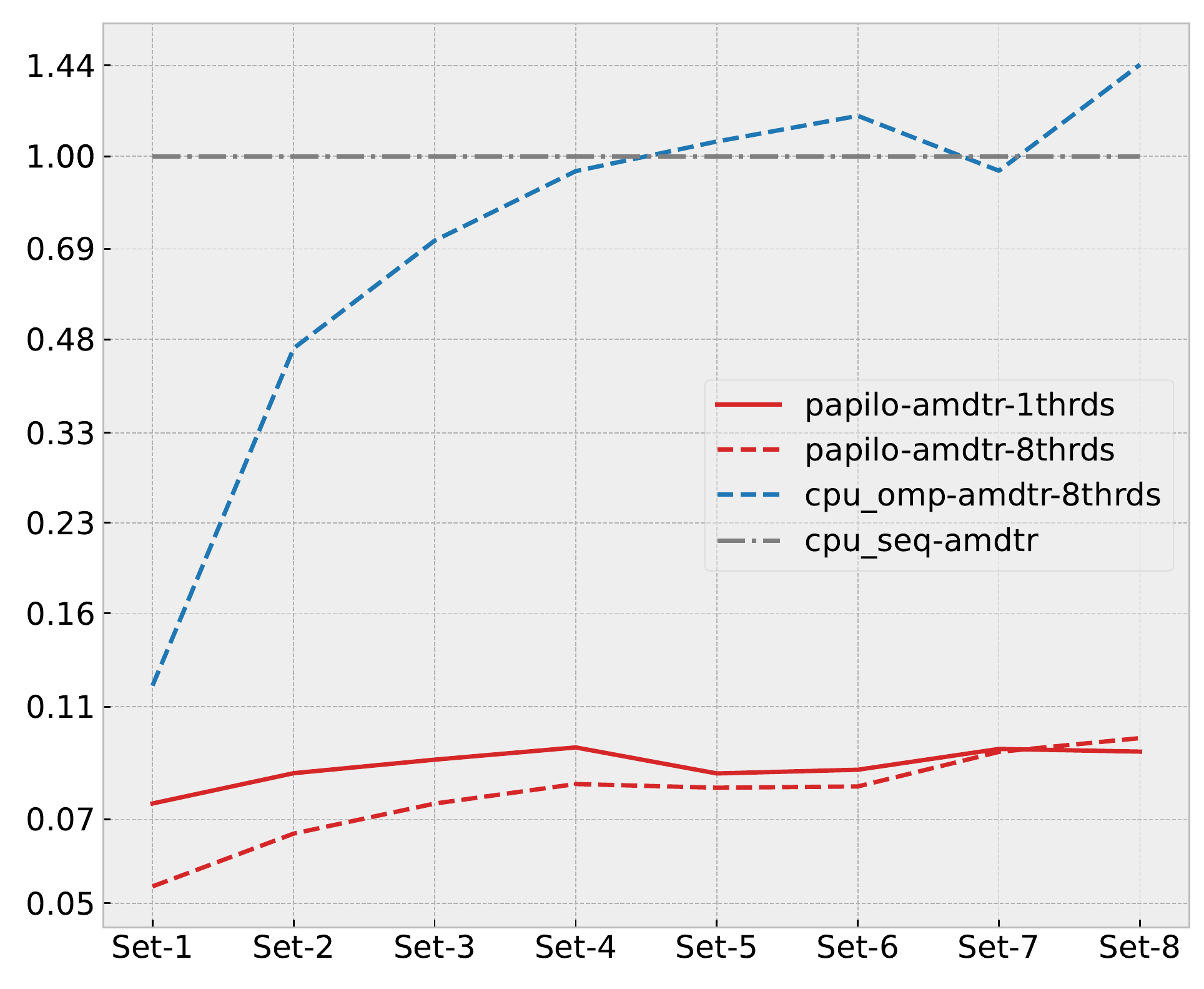}}
  %\vspace{-1.0em}
\caption{Geometric means of speedups over the eight
  subsets of instances of increasing size for the two PaPILO executions with \num{1} and \num{8} threads, and the \cpuomp execution. All executions are in double-precision arithmetic. Baseline is \cpuseq running on the \amdtr machine.}
\label{fig:papilo}
\end{figure}

From the point of view of correctness and comparison of results, the above-mentioned behavior of PaPILO also presents difficulties. Because PaPILO will often reduce the problem formulation as described, we can not directly compare the arrays of tightened bounds. Instead, we run our code and record the runtime, but afterwards, we pass the already propagated instance to PaPILO and let it process it. PaPILO will perform the necessary reductions such that the problem then (hopefully) matches the problem that is solved by PaPILO from scratch. However, apart from checking that the resulting tightened bounds are the same as described in Section \ref{sub:comparison_metric}, we now also check that PaPILO did not find any bound changes in the instance solved by our code first. We acknowledge that this is not a rigorous way to check if the results are the same, but we believe it suffices for our goal of establishing that our baseline implementations are valid benchmarks in terms of accuracy and performance. Additionally, this approach re-reads and re-writes instances multiple times through two different MIP instance file readers, which in few cases results in errors and wrongly read values. These and other issues reduced the number of instances where PaPILO and our code produce the same results. Nevertheless, \num{701} instances did pass this check and they form the base for the results shown in Figure \ref{fig:papilo}.

All executions are done on the \amdtr machine (see Section \ref{sub:hardware}) in double-precision arithmetic. The base case is \cpuseq running on the \amdtr machine against which the speedups are computed for the two PaPILO executions with \num{1} and \num{8} threads, and \cpuomp execution with \num{8} threads. On average, the single-threaded PaPILO run achieves a speedup of \num{0.08} compared to our single-threaded code. The PaPILO run with \num{8} threads performs sightly worse with a speedup of \num{0.07}, on average. As is the case with our code, the multi-threaded PaPILO execution performs worse for subsets with smaller instances, while it eventually beats the single-threaded execution for the \setn{8} with the largest instances.
All in all, these results demonstrate that our implementations \cpuseq and \cpuomp provide a competitive baseline for the performance comparisons presented in the previous sections.

\section{Conclusions and Outlook}

In this paper, we show that domain propagation is --- with the right algorithmic design --- amenable to efficient execution on GPUs. While this result is promising for the exploitation of GPUs in the MIP solving process, challenges still remain in this respect. Two primary use cases for domain propagation in MIP solvers are in \emph{presolving} and after \emph{branching} on variables during the main solving process. In the former case, the algorithm starts in the same setting as used in this paper - with no prior knowledge of the state of the constraint system. In the latter case, however, the system was usually already fully propagated before the branching took place. This means that the algorithm starts at a situation that is equivalent to just after a propagation round with a single bound change on the branching variable. Here, the \cpuseq can in most cases avail of the constraint marking mechanism to significantly reduce the amount of work it needs to perform (see Section \ref{sub:gpuparallelalg}), to the point where there is not enough work to justify the cost of parallelization. On the other hand, the amount of time spent in domain propagation during presolving is usually small relative to the main solving time, and thus the motivation to speed it up  ``as is'' is low. 

In conclusion, while our new algorithm offers the ability to perform domain propagation on a much larger scale than before, new parent methods are required which will use domain propagation in a way suited for GPUs and to the benefit of MIP solvers. A particularly useful feature of our algorithm in this respect is that it is able to run without the need for synchronization with the CPU, making it possible to embed it in parent GPU-based algorithms as well as allowing the CPU to keep processing while the GPU propagation is taking place.

Execution of SpMVs on GPUs is an active research area and several improvements on the CSR-adaptive from \cite{GreathouseDaga2014} as well as new SpMV algorithms have been published since. For example, see \cite{Daga2015StructuralAS,glob_homogen_spmv,Li2021AdaptiveSO}. A future research direction would be to analyze if the improvements of SpMV algorithms in the recent literature can be carried over to the computation of activities as described in this paper (see Section \ref{sec:impl}).

\bibliographystyle{abbrv}
\bibliography{arxiv_main}

\appendix
\clearpage
\appendixpage

\section{Variability of Baselines}

\begin{figure}[htbp]
  \centerline{\includegraphics[width=0.5\linewidth]{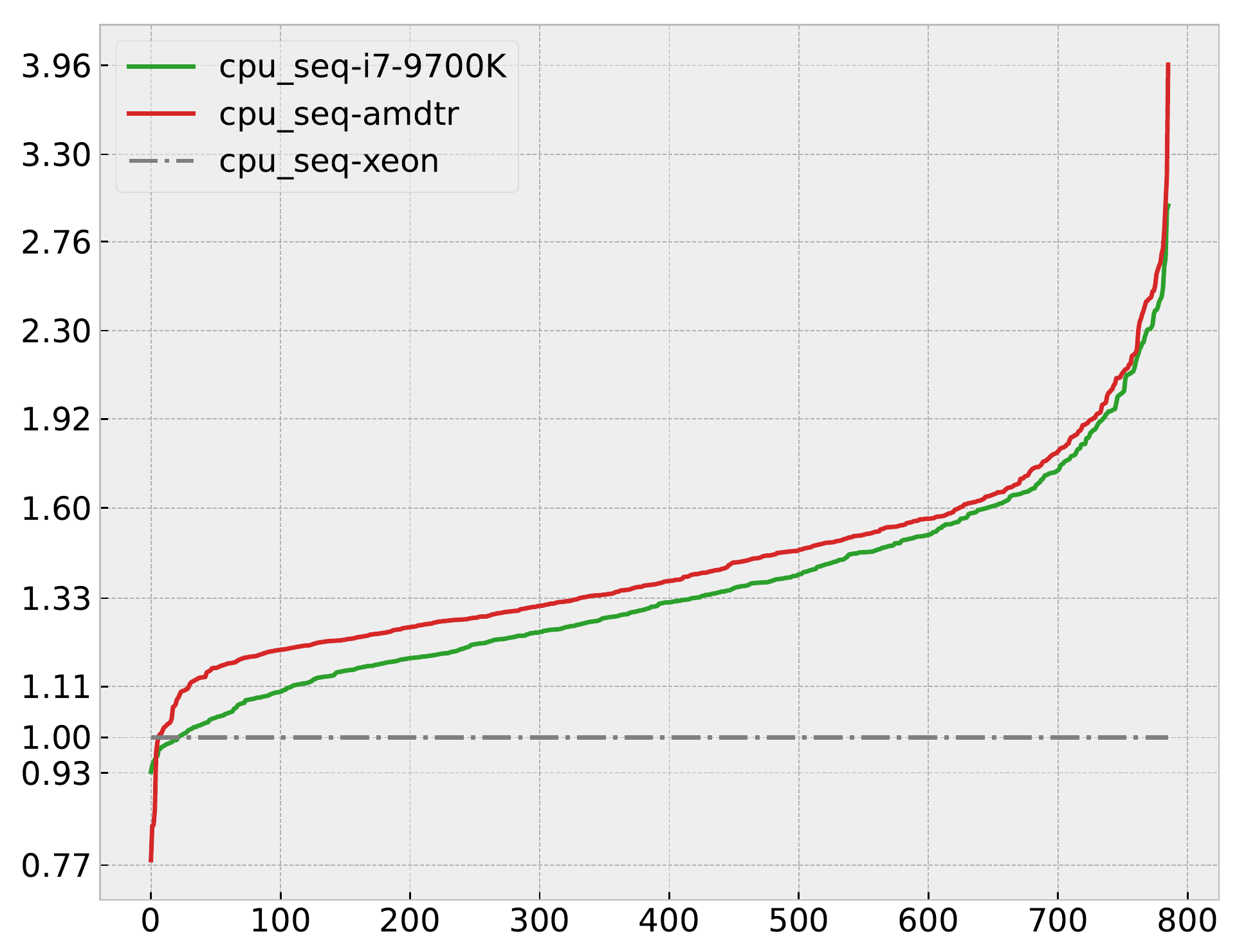}}
  %\vspace{-1.0em}
\caption{Distribution of speedups by plotting the individual speedup for each instance,
  sorted in ascending order. All executions are in double-precision arithmetic. Baseline is \cpuseq running on the \xeon machine.}
\label{fig:baselines_double}
\end{figure}

The CPU architectures where we execute the \cpuseq algorithm differ considerably among themselves (see Section \ref{sub:hardware}). As explained, we chose the execution of \cpuseq on the \xeon machine as our baseline case and reported all speedups relative to this execution. However, one should keep in mind that executions of the \cpuseq algorithm on other architectures will give different results due to architectural differences (e.g., amount of RAM or cache memory). This effect is shown in Figure \ref{fig:baselines_double}, which plots the speedups of the \cpuseq execution on \amdtr and \planck machines relative to the execution on the \xeon machine on the instances from the test set defined in Section \ref{sub:testset}. We can see that the \amdtr execution is faster on average than the \xeon run, but not by a constant factor. In fact, there are instances where the \amdtr is almost \num{4} times faster, and other instances where \xeon beats it by a factor of \num{1.2}. This shows that as the features of the instances change (e.g., size, shape of constraint matrix, etc.), the architecture of the machine affects performance by a different amount. Furthermore, the slopes on the curves are not linear, indicating that for some instances, the architectural differences have a disproportionately high effect.

\section{Effect of Constraint and Variable Ordering on Performance}
\label{ap:ordering}

\begin{figure}[htbp]
  \centerline{\includegraphics[width=0.5\linewidth]{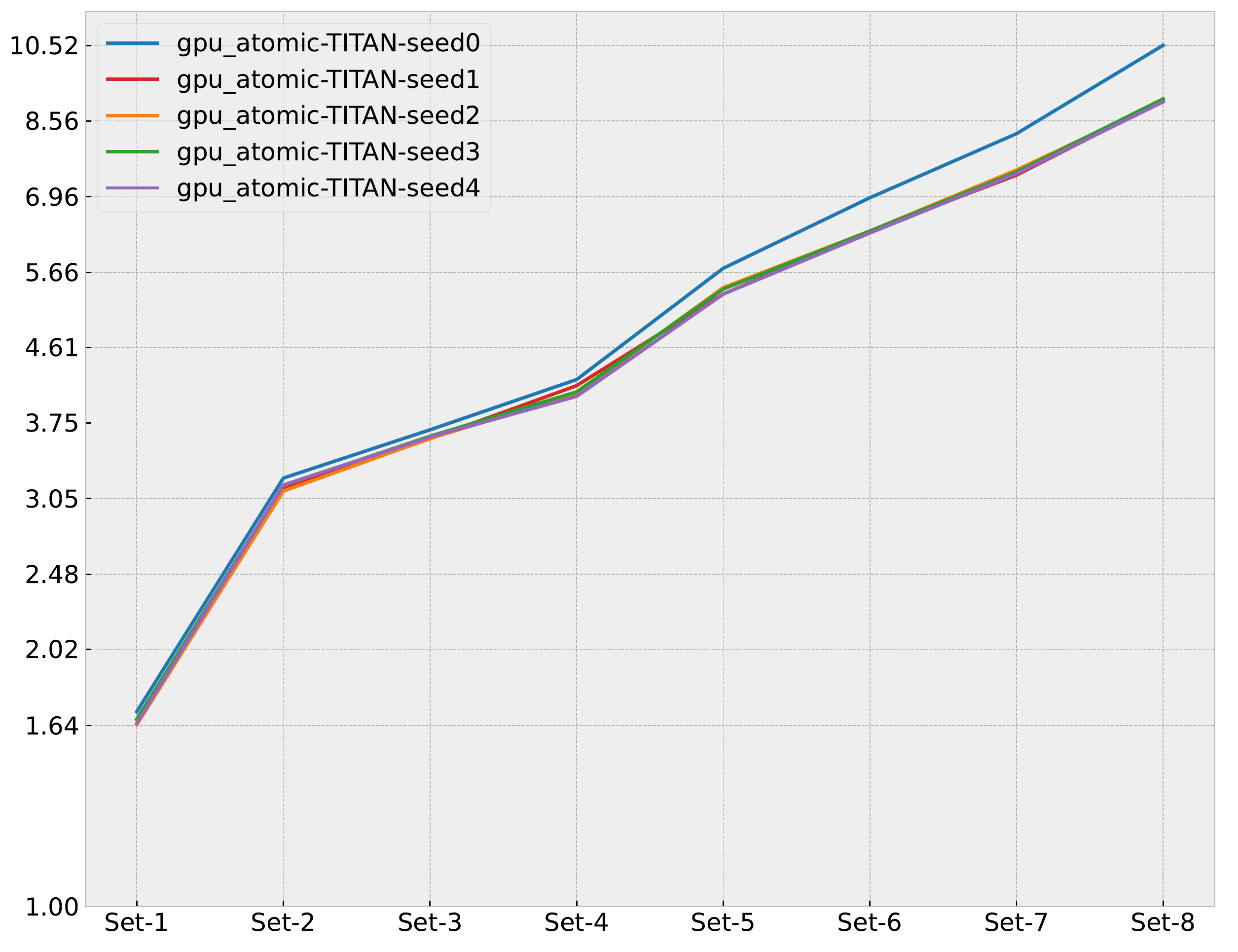}}
\caption{Geometric means of speedups over the eight subsets of instances of increasing size. Each line belongs to the execution with a different ordering of variables and constraints in the test set. Ordering \textit{seed0} is the default ordering of the original test set. All executions are in double-precision arithmetic. Baseline is \cpuseq running on the \xeon machine with the default ordering.}
\label{fig:seeded}
\end{figure}

The possibility to perform a specific bound change can depend on other bound changes being applied first. This means that depending on the order in which the constraints are processed, propagation algorithms can in principle alter their performance. For example, consider the cascading propagation example introduced in Section \ref{sub:reduced_eff}. If the sequential algorithm processes these constraints in order from the first towards the last, it can propagate this system in one round. On the other hand, if the processing happens in the reverse order, from the last towards the first constraint, it would need $m$ rounds, with $m$ being the number of constraints. In this appendix, we evaluate to what extent the effect of ordering of constraints and variables influences the results presented in Section \ref{sec:compres}. We run the experiments from Section \ref{sub:results} on the \titan machine, but this time on instances with randomly permuted ordering of constraints and variables, as well as on the instances with the original ordering. The results are shown in Figure \ref{fig:seeded}.      

As we can see, the difference in speedups among the runs with randomly permuted input is negligible. Run \textit{seed0} with the original ordering performs better than the other runs, but not by a significant amount: averaged over the whole test set, the difference in speedups between them does not exceed \num{4.3} percent. While this difference in behavior of the run \textit{seed0} is interesting, it is not surprising, as the original ordering is not random but rather made by hand, where similar constraints are often grouped together.

\section{GPU Versus CPU Synchronization}
\label{ap:gpucpuloop} 

\begin{figure}[htbp]
  \centerline{\includegraphics[width=0.5\linewidth]{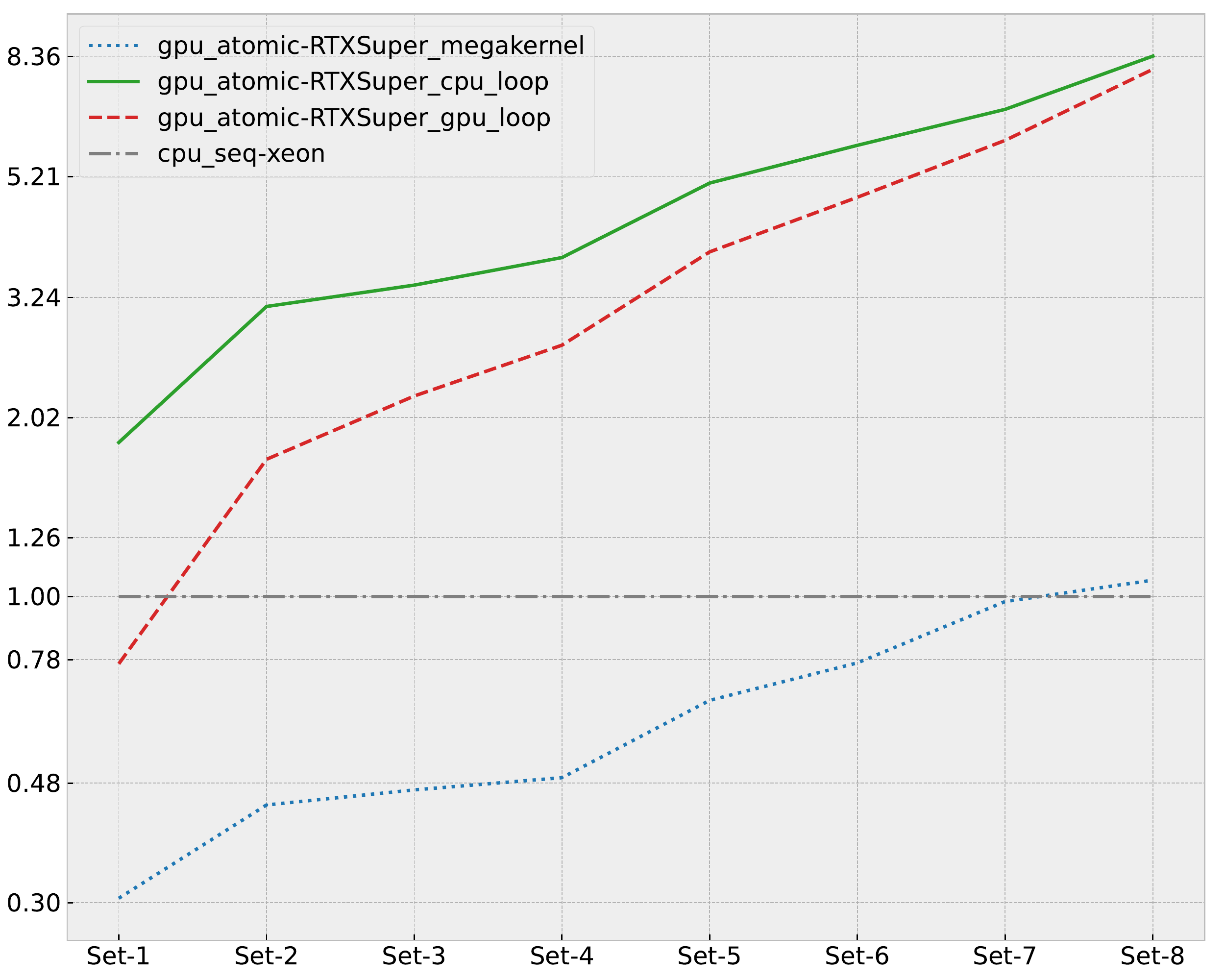}}
  %\vspace{-1.0em}
\caption{Geometric means of speedups over the eight
  subsets of instances of increasing size for the three algorithm variants \cpuloop, \gpuloop, and \megaker. Baseline is \cpuseq running on the \xeon machine. All executions are in double-precision arithmetic.}
\label{fig:cpugpuloop}
\end{figure}

In Section \ref{sub:cpugpuloop} we presented three variants of kernel synchronization: \cpuloop, \gpuloop, and \megaker. We now present results quantifying the difference in their performance. The test set as defined in Section \ref{sub:testset} is run on the \planck CPU and the \super GPU (see Section \ref{sub:hardware}). Results are shown in Figure \ref{fig:cpugpuloop}.

Over the whole test set, the \cpuloop is \num{1.72} times faster than the \gpuloop, on average. However, most of this speedup comes from instances of smaller size, as we can see in Figure \ref{fig:cpugpuloop}. The sequentialization point, which causes the slowdown on the GPU, remains constant as the sizes of instances increase, while the parallel part exploited by the GPU increases. Hence, as the sizes of the instances increase, the sequentialization point has less and less effect on the performance, an effect described in Amdahl's law \cite{10.1145/1465482.1465560}. Indeed, we can see in Figure \ref{fig:cpugpuloop} that the two curves converge as the problem sizes increase.

On the other hand, the \megaker performs significantly worse than the other two variants over all subsets. Finally, keep in mind that our implementations were developed for the purpose of this paper, and while they contain all the described algorithmic steps, numerous performance optimizations are still possible. Due to the higher code complexity of the \megaker compared to the \cpuloop and \gpuloop, it could perhaps benefit the most from future optimization efforts.   

\section{Differences to the Conference Version}
A conference version of this paper was submitted and accepted for publication in the proceedings of IEEE/ACM 10th Workshop on Irregular Applications: Architectures and Algorithms (IA3) \cite{SofranacGleixnerPokutta2020}.
%In this appendix, we highlight the most important differences and improvements made in the meantime.
Since then, the code has been significantly extended and improved, including bug fixes. Some of these additions are performance-neutral, while others affected it negatively. Most notably, we acknowledge a missing compiler flag during our test execution which negatively affected the performance of the \cpuseq baseline used in \cite{SofranacGleixnerPokutta2020}. On the other hand, many new additions improve performance, e.g., the reduced number of atomic operations described in Section \ref{sub:bdcands}, the possibility to run the rounds loop on the CPU described in Section \ref{sub:cpugpuloop}, technical code improvements to reduce register usage and others. All these changes affect performance in a non-uniform way: in some cases, the performance is now superior to the results reported in \cite{SofranacGleixnerPokutta2020}, while in other cases it is inferior. 

Also, the correctness of results was affected in some cases due to a missing feature described in Section \ref{sub:infcontr}. Moreover, numerical handling has been significantly improved, making the code more robust against round-off errors and other numerical difficulties. However, note that also in \cite{SofranacGleixnerPokutta2020}, performance comparisons were only reported over sets of instances for which competing implementations had converged to the same result.
  
\end{document}